\documentclass[multphys,vecphys]{svmult}
\usepackage{makeidx} 
\usepackage{graphicx}
\usepackage{multicol} 
\usepackage[bottom]{footmisc}

\makeindex  

\begin{document}

\title*{Theoretical Interpretation of GRB 031203 and URCA-3.}
\author{Remo Ruffini\inst{1,2}, Maria Grazia Bernardini\inst{1,2}, Carlo Luciano Bianco\inst{1,2}, Pascal Chardonnet\inst{1,3}, Federico Fraschetti\inst{1} \and She-Sheng Xue\inst{1,2}}
\authorrunning{Ruffini et al.}
\institute{ICRANet and ICRA, P.le della Repubblica 10, I--65100 Pescara, Italy.
\texttt{ruffini@icra.it}
\and Dip. Fisica, Univ. ``La Sapienza'', P.le A. Moro 5, I--00185 Roma, Italy.
\and Univ. de Savoie, LAPTH--LAPP, BP 110, F--74941 Annecy, France.}

\maketitle

\section{Luminosity and spectral properties.}
GRB 031203 was observed by IBIS, on board of the INTEGRAL satellite in the $20-200$ keV band \cite{mg}, as well as by XMM \cite{wats} and Chandra \cite{soderb} in the $2-10$ keV band, and by VLT \cite{soderb} in the radio band. It appears as a typical long burst \cite{saz}, with a simple profile and a duration of $\approx 40$ s. The burst fluence in the $20-200$ keV band is $(2.0\pm 0.4)\times 10^{-6}$ erg/cm$^2$ \cite{saz}, and the measured redshift is $z=0.106$ \cite{proch}. We analyze in the following the gamma-ray signal received by INTEGRAL.

Such observations find a direct explanation in our theoretical model \cite{rubr}. We determine the values of its two free parameters: the total energy stored in the Dyadosphere $E_{dya}$ and the mass of the baryons left by the collapse $M_Bc^2=BE_{dya}$. We follow the expansion of the pulse, composed by the $e^--e^+$ plasma initially created by the vacuum polarization process in the Dyadosphere. The plasma self--propels outward and engulfs the baryonic remnant left over by the collapse of the progenitor star. As such pulse reaches transparency, the Proper--GRB is emitted. The remaining accelerated baryons interacting with the ISM produce the afterglow emission. The ISM is described by the two additional parameters of the theory: the average particle number density $n_{ISM}$ and the ratio $\mathcal{R}$ between the effective emitting area and the total area of the pulse \cite{spectr1}, which takes into account the ISM filamentary structure \cite{spectr2}. The best fit of the observational data (see Fig.\ref{ruffiniF2}.Left) leads to a total energy of the Dyadosphere $E_{dya}=1.85\times10^{50}$ erg and the amount of baryonic matter in the remnant is $B = 7.4\times10^{-3}$. The ISM parameters are: $<n_{ism}>=0.3$ particle/$cm^3$ and $<\mathcal{R}>=7.81\times 10^{-9}$ \cite{031203}.
\begin{figure}
\centering
\includegraphics[width=0.49\hsize,clip]{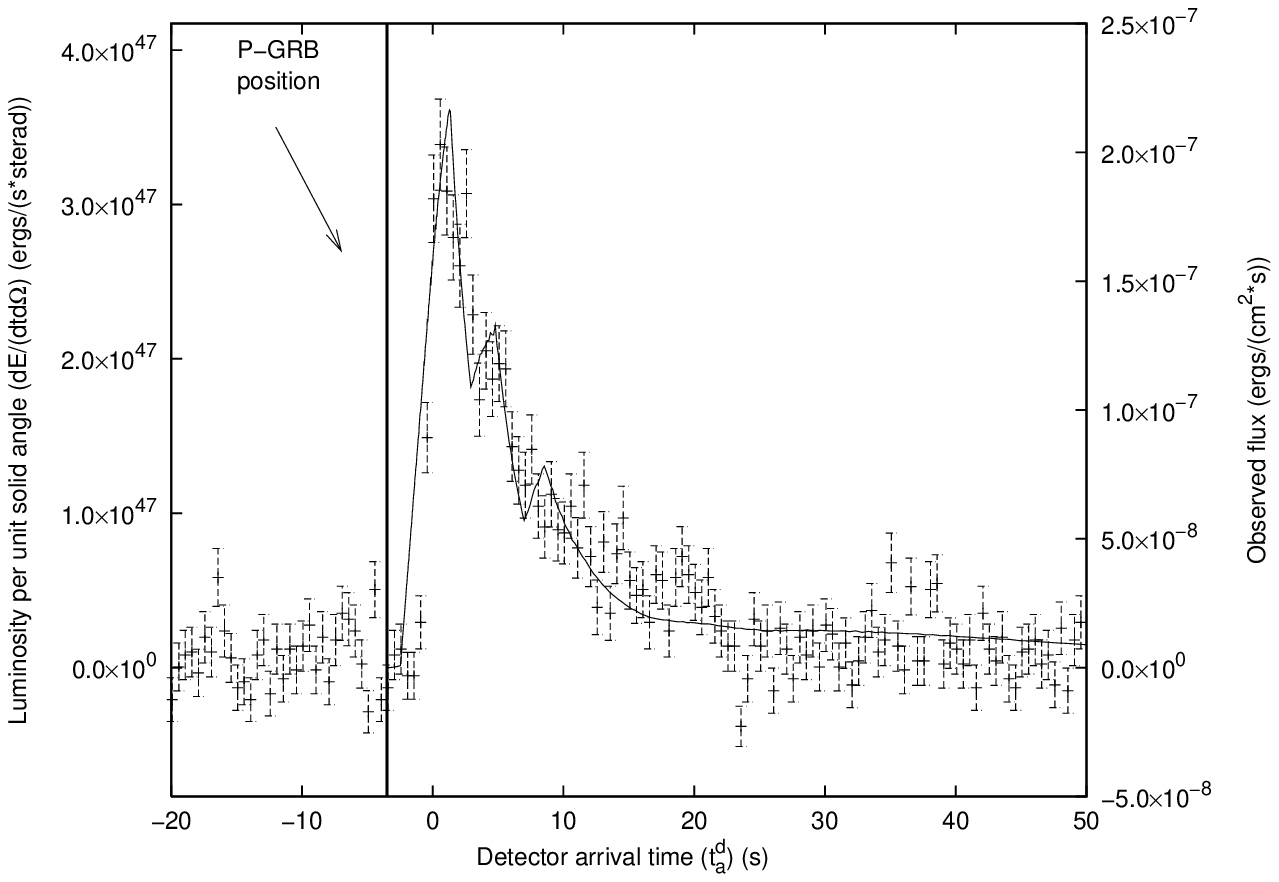}
\includegraphics[width=0.49\hsize,clip]{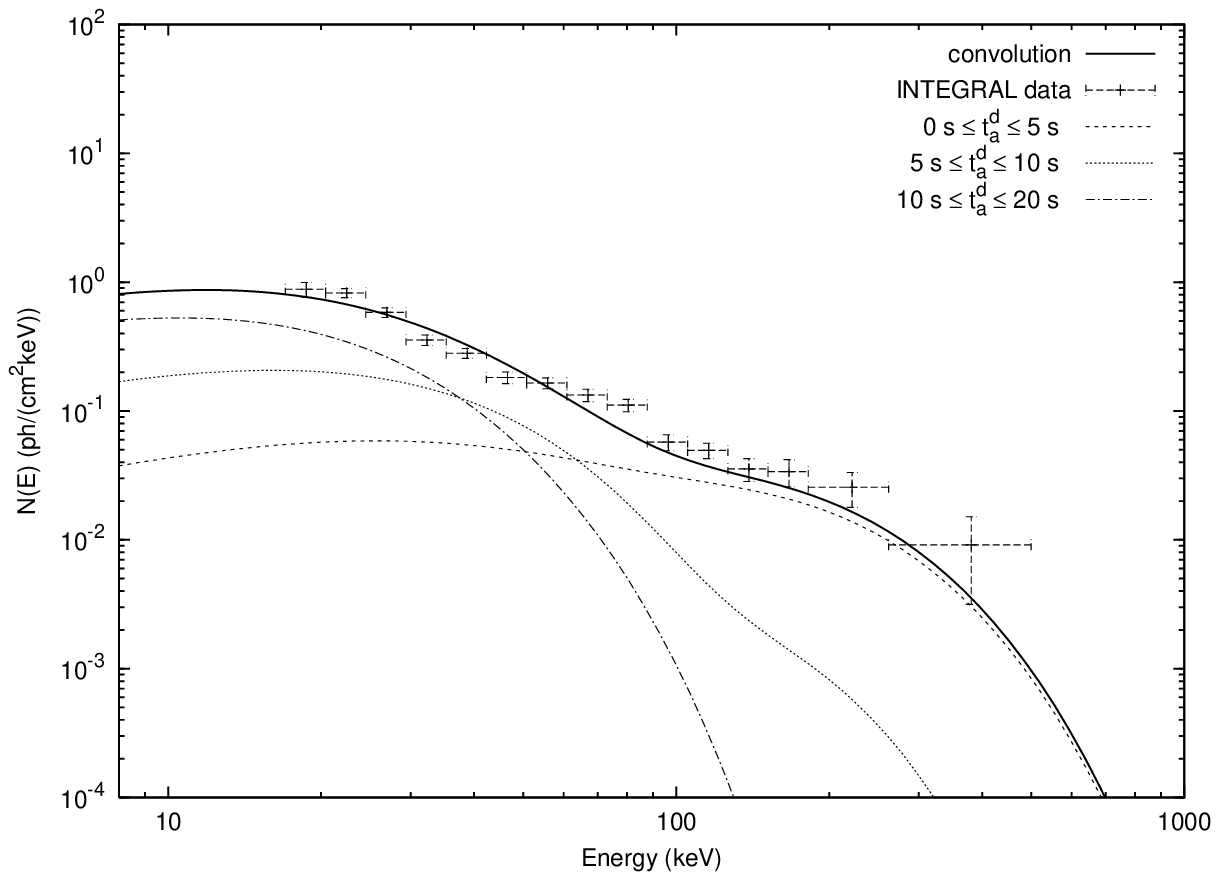}
\caption{\textbf{Left:} Theoretically simulated light curve of the GRB 031203 prompt emission in the $20-200$ keV energy band (solid line) is compared with the observed data \cite{saz}. \textbf{Right:} Three theoretically predicted time--integrated photon number spectra $N(E)$ are here represented (dashed, dotted, dashed--dotted lines). The theoretically predicted time--integrated photon number spectrum $N(E)$ corresponding to the first $20$ s of the prompt emission (solid line) is compared with the data observed by INTEGRAL \cite{saz}.}
\label{ruffiniF1}
\end{figure}

The luminosity in selected energy bands is evaluated integrating over the EQuiTemporal Surfaces (EQTS) \cite{eqts} the energy density released in the interaction of the accelerated baryons with the ISM measured in the co--moving frame, duly boosted in the observer frame. The radiation viewed in the co--moving frame of the accelerated baryonic matter is assumed to have a thermal spectrum \cite{spectr1}. In addition to the luminosity in fixed energy bands we can derive also the instantaneous photon number spectrum $N(E)$. Although the spectrum in the co--moving frame of the expanding pulse is thermal, the shape of the final spectrum in the laboratory frame is non thermal. In fact each single instantaneous spectrum is the result of an integration of hundreds of thermal spectra over the corresponding EQTS \cite{eqts}. This calculation produces a non thermal instantaneous spectrum in the observer frame. We integrated the photon number spectrum $N(E)$ over the whole duration of the prompt event: in this way we obtain a typical non--thermal power--law spectrum which results to be in good agreement with the INTEGRAL data and gives a clear evidence of the possibility that the observed GRBs spectra are originated from a thermal emission \cite{031203} (see Fig.\ref{ruffiniF1}.Right).

\section{The GRB 031203/Sn2003lw/URCA-3 connection.}
In the early days of neutron star physics it was clearly shown by Gamow and Shoenberg \cite{gs} that the URCA processes are at the very heart of the Supernova explosions. The neutrino--antineutrino emission described in the URCA process is the essential cooling mechanism necessary for the occurrence of the process of gravitational collapse of the imploding core. Since then, it has become clear that the newly formed neutron star can be still significantly hot and in its early stages will be associated to three major radiating processes \cite{tsu}: a) the thermal radiation from the surface, b) the radiation due to neutrino, kaon, pion cooling, and c) the possible influence in both these processes of the superfluid nature of the supra-nuclear density neutron gas. 

\begin{figure}
\centering
\includegraphics[width=0.55\hsize,clip]{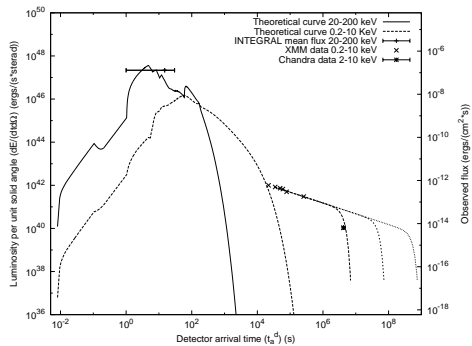}
\caption{Theoretically simulated light curves of GRB 031203 in $20-200$ KeV (solid line) and $2-10$ KeV (dashed line) energy bands are represented together with qualitative representative curves for the cooling processes.}
\label{ruffiniF2}
\end{figure}

We already proposed that the Supernova explosion can be the result of an induced gravitational collapse \cite{lett3}, and the X--ray emission observed is the sign of the cooling of the young neutron star born from the Supernova explosion (see GRB 980425/SN1998bw/URCA-1 \cite{cospar} and GRB 030329/SN2003dh/URCA-2 \cite{030329}). This possibility was explored also for the system GRB 031203/SN2003lw. Also in this case the observations in other wavelengths could be related to the Supernova event, which we called URCA-3 (see Fig.\ref{ruffiniF2}).

\end{document}